\documentclass[runningheads,a4paper]{llncs}
\usepackage{fancyvrb}
\usepackage{url}
\usepackage{amsmath}
\usepackage{amssymb}
\usepackage{listings}
\usepackage{graphicx}
\usepackage{hyperref}
\usepackage{bookmark}
\usepackage[margin=1.1in]{geometry}
\hypersetup{hidelinks}
\newcommand{\keywords}[1]{\par\addvspace\baselineskip
\noindent\keywordname\enspace\ignorespaces#1}

\begin{document}
\lstset{basicstyle=\ttfamily\small,commentstyle=\ttfamily\small,
tabsize=2,breaklines,showstringspaces=false}

\mainmatter

\title{Evoking Harmony via Convolution%
\texorpdfstring{\thanks{Extended version of a paper
to appear in the Proceedings of the 8th International Csound Conference
(ICSC 2026), Trapani, 20--22 October 2026.}}{}}
\titlerunning{Evoking Harmony}

\author{Michael Gogins\inst{1}}
\authorrunning{M.~Gogins}
\institute{Irreducible Productions, New York\\
\url{http://michaelgogins.tumblr.com}\\
\email{michael.gogins@gmail.com}}

\maketitle

\begin{abstract}
I show how to evoke the pitch-class content of a chord from an arbitrary
source sound by convolving the source with an impulse response whose
grains are one windowed sinusoid per pitch-class, across each octave of hearing range; while, at the same time, minimizing artifacts.
A Csound user-defined opcode (\texttt{chord\_convolver}) mixes a dry Dirac component into that
response, and applies partitioned convolution once. I contrast the effect with
a linear-frequency comb filter and with a generic constant-$Q$ resonator bank,
and I demonstrate musical use on a twilight field recording alongside the
ruins of Ch\^{a}teau de Lagarde.
\keywords{csound, harmony, convolution, constant-$Q$, field recording}
\end{abstract}

\section{Introduction}

When I was a boy my father drove with me from Salt Lake City to Los Angeles
across the desert at night. Long stretches of highway were empty and straight.
I lay with my head against the window and listened to the car, the road, and
the wind. In that rushing noise I heard melodies and chords---tonality arising
from broadband sound.

This paper reports experiments in Csound~\cite{csound_home} that evoke
such harmonies from varied sources. Conceptually, the effect convolves a
source signal with an impulse response whose grains occupy only the pitch-classes of
a chosen chord, realized as sinusoids across every octave within hearing range.
I call the result \emph{evoking harmony} (or harmony convolution). Materials
are available online~\cite{gogins_github_io}.

The effect sits at the boundary of timbre and tonality. A constant-$Q$
filterbank centered on the same pitch-classes is closely related; a generic
constant-$Q$ tiling that is \emph{not} restricted to a chord is only loosely
related; a classical comb filter is even less related, because its teeth are
periodic in \emph{linear} Hz rather than in log frequency. Gabor- or
Morlet-style wavelet cells, not being centered on pitch-classes, are not a
good fit.

\section{Motivation}

There has long been a practical disjunction between ``non-tonal''
electroacoustic practice---often oriented toward acousmata and continuum of
timbre---and tonal practice oriented toward pitch relations abstracted from
actual sounds. That split goes back to the roots of computer music in the
1950s and 1960s, a period dominated by technological utopianism and musical
serialism. The algorithm described here deliberately works on the boundary of
timbre and tonality: it leaves the source audible while imparting a faint or
strong sense of chords or chord progressions, depending on parameters and mix.

\section{Mathematics}

Let $C$ be a finite set of pitch-classes. A \emph{grain} is one causal
half-cosine windowed sinusoid (pitch-class $c$, octave $n$). Its frequency
response is a \emph{kernel} $H_{c,n}$. The \emph{impulse response} is the sum
of the loudness-balanced grains plus a Dirac, then jointly $\ell_2$-normalized.
For a chord $C$ those kernels form an octave-related scaled family:
\[
H_C(f)
=
\sum_{c\in C}\sum_n H_{c,n}(f),
\qquad
H_{c,n}(f)
\simeq
H_c\!\left(2^{-n}f\right).
\]
In log-frequency coordinates $u=\log_2 f$, octave transposition is translation
by one unit, and the total response is approximately periodic with period one
octave:
\[
H_C(u+1)\simeq H_C(u).
\]
Each octave therefore contains $K=|C|$ peaks whose positions within the octave
are fixed by the chord. The chord acts as the unit cell of a periodic filter
in log-frequency space. That is the structural difference from a comb filter,
whose peaks are equally spaced in linear Hz.

Convolution with a finite impulse response incurs Gabor's time--frequency
tradeoff~\cite{gabor1946theory}: a shorter grain improves time resolution at
the cost of frequency resolution, while a longer grain improves frequency resolution 
at the cost of time resolution. 
The duration of the grain is therefore a musical parameter. The
frequency-dependent window length $T(f)=\min(T,\,T\cdot 440/f)$ used below
distributes that tradeoff unevenly across the lattice, and deliberately so.
Above a pivot at $440\,\mathrm{Hz}$, $T(f)=T\cdot 440/f$ holds the number of
cycles in every grain constant at $T\cdot 440$, so the treatment is
constant-$Q$ with $Q=T\cdot 440$: temporal smear falls as $1/f$, while spectral
smear stays constant in the very coordinate $u=\log_2 f$ in which the chord is
the unit cell. For $T=30\,\mathrm{ms}$ the measured $-3\,\mathrm{dB}$ bandwidth
is $1.68$ semitones in every octave from $440\,\mathrm{Hz}$ upwards. Below the
pivot the length is capped at $T$, so bandwidth is constant in Hz instead---some
$43\,\mathrm{Hz}$ for the same $T$---and therefore widens in musical terms as
frequency falls, to $3.4$ semitones at $220\,\mathrm{Hz}$ and $14.6$ at
$55\,\mathrm{Hz}$. Smearing is thus least, and the transients of the source best
preserved, in the upper octaves; it is the bass that rings and blurs.

There is a further reason to hold the grains short, and it is perceptual rather
than spectral. The aim is that the harmony seem latent in the source and evoked from
it, not added to it. Convolution serves that aim by construction,
since $Y(f)=H(f)X(f)$ vanishes wherever the source has no energy: convolution 
can only expose pitch content that the source already contains, never introduce any.
What can still betray the effect is time. Common onset binds concurrent sounds
into a single event, but energy persisting after the source has fallen silent
would be a cue for a separate, struck resonator -- not a property of the
source. The causal window and the cap on $T(f)$ both serve auditory fusion in
this sense, by keeping the harmonic response within the duration of the source.
On a $50\,\mathrm{ms}$ impulse response driven by a $3\,\mathrm{Hz}$ impulse
train, the design used here leaves under $3\,\%$ of each period's energy
outside the interval the dry source occupies, whereas an uncapped or
symmetrically windowed grain displaces a third or more. On sustained material 
the tail is never exposed and longer grains
cost nothing, which makes $T$ properly a function of the transient density of
the source rather than a constant of the method.

\section{Method}

\subsection{Implementation}

The Csound user-defined opcode is
\texttt{chord\_convolver}. It builds one impulse response $h$, and
applies partitioned convolution once via \texttt{ftconv}~\cite{csoundmanualftconv}.
The design choices are:

\begin{enumerate}
\item
  For each MIDI note whose pitch-class is selected, include that frequency if
  it lies between $20\,\mathrm{Hz}$ and $0.95$ of Nyquist.
\item
  Give partial $f$ a causal half-cosine grain of length
  \[
  T(f)=\min\bigl(T,\,T\cdot 440/f\bigr),
  \]
  keeping ringing equal across higher partials. A full Hann peaking mid-window would delay the
  audible onset of the grain, loosening its fusion with the attack of the
  source; the half-cosine starts at $1$ and fades to $0$.
\item
  Scale each partial by $A(f)\propto T(f)^{-1/2}$, so that all partials carry
  equal energy despite their differing lengths (Section~\ref{sec:levelling}).
\item
  Weight each partial by inverse IEC A-weighting (unity at $1\,\mathrm{kHz}$),
  with a rumble shelf that evaluates $R_A$ no lower than $150\,\mathrm{Hz}$ and
  a boost cap of $+6\,\mathrm{dB}$, so that low wind and rumble do not dominate.
\item
  Scale grains by an impulse gain, average across partials, add a Dirac
  component at sample $0$ (dry path inside the IR), apply a joint $\ell_2$
  normalization to the whole IR, and convolve.
\end{enumerate}

\subsection{Levelling Octaves}
\label{sec:levelling}

Because $T(f)$ shrinks with frequency, partials of equal amplitude are not
equally present. Levelling them proceeds in two stages: first equalize the
physical contribution of each partial, then convert that into equal perceived
loudness.

For the first stage the relevant quantity is grain energy, not amplitude and not
the window integral. The sources of interest are broadband, so for excitation of
power spectral density $S_0$ the output power contributed by partial $k$ is, by
Parseval's theorem,
\[
P_k \;=\; S_0\!\int \bigl|G_k(f)\bigr|^2 df
     \;=\; S_0\!\int g_k(t)^2\,dt
     \;=\; S_0 E_k ,
\]
so equal presence means equal $E_k$, giving $A(f)\propto T(f)^{-1/2}$. A
constant factor then restores the total grain energy, so that the impulse and
Dirac gains retain their meaning independently of this choice.

Loudness tracks that power rather than spectral density because every grain is
narrower than one auditory filter: measured against the auditory equivalent rectangular bandwidth (ERB) 
of~\cite{glasberg1990derivation}, the widest grain in the lattice occupies
$0.74$ of a critical band and the rest fall below $0.5$. This is why the window
integral $A(f)T(f)$, which fixes the peak transfer gain $|H(f_k)|$, is the wrong
target: as $T(f)$ shortens, the grain's bandwidth $\approx 1/T(f)$ widens, and
holding peak gain flat across a widening band yields rising power. Normalizing
the integral instead of the energy measurably tilts the lattice bright by some
$8\,\mathrm{dB}$ from the bass to the top octave.

The second stage is the inverse A-weighting of item~4 above, which converts
equal power per partial into approximately equal loudness per partial. Measuring
the realized impulse response confirms the two stages are cleanly separated: the
per-partial band powers follow the A-weighting curve to within
$0.02\,\mathrm{dB}$ across the lattice, so nothing but the intended perceptual
weighting remains.

Equal loudness \emph{per partial} is not the same as equal loudness \emph{per
critical band}. Since an ERB spans some 27 semitones at $20\,\mathrm{Hz}$ but
only 1.9 semitones above $4\,\mathrm{kHz}$, the low octaves crowd several chord
tones into one auditory filter, where their powers sum. Dividing out that
crowding is available as an option, and attenuates the lowest octaves by 4 to
$6\,\mathrm{dB}$; the recordings here do not use it, keeping each lattice tone
reading equally rather than evening out the harmony as a whole.

Pseudocode:

\begin{verbatim}
chord_convolver(x, T, g_imp, g_dirac, compensation, PCs...):
  T     = clamp(abs(T), 0.01, 1.5)
  n_max = T * 440
  N     = max(2, round(T * sr))
  // scale degrees reduce to pitch-classes; negatives are unused
  PCs   = { p mod 12 : p in PCs, p >= 0 }
  freqs = [];  weights = [];  Tf = [];  amp = []
  for midi = 0..127:
    if (midi mod 12) not in PCs: continue
    f = cpsmidinn(midi)
    if f < 20 or f > 0.95*Nyquist: continue
    fp = max(f, 150)            // rumble shelf
    w  = R_A(1000) / R_A(fp)    // inverse A-weight
    w  = min(w, 10^(6/20))      // cap boost +6 dB
    append f to freqs;  append w to weights
    append min(T, n_max / f) to Tf
  M = length(freqs)
  p = [0, 0.5, 1, 0.5][compensation]  // 3 also * D_k^-1/2
  for i = 0..M-1:  amp[i] = (T / Tf[i])^p
  c = sqrt(sum(Tf) / sum(amp^2 * Tf))  // hold total energy
  for i = 0..M-1:  amp[i] = c * amp[i]
  h = zeros(N)
  for n = 0..N-1:
    t = n / sr;  s = 0
    for i = 0..M-1:
      if t < Tf[i]:
        env = 0.5 + 0.5*cos(pi*t/Tf[i])   // causal, 1 -> 0
        s = s + weights[i]*amp[i]*env*sin(2*pi*freqs[i]*t)
    h[n] = (s / M) * g_imp
  h[0] = h[0] + g_dirac
  h    = h / L2norm(h)
  return ftconv(x, h)
\end{verbatim}

Earlier experiments compared IIR resonator banks and multi-\texttt{ftconv}
variants; the joint IR of \texttt{chord\_convolver} is the implementation used
in the comparison spectrograms and the sound walk below.

\subsection{What It Is Not}

A \emph{constant-$Q$} bank with centers restricted to the chord's pitch-classes
is a near neighbour: same log lattice, different dynamics (ongoing filtering
versus convolution with a finite impulse response). A generic constant-$Q$ tiling spaced
every few semitones (not chord-restricted) emphasizes a regular pitch grid
without chord identity. A \emph{comb} filter with delay $1/f_0$ emphasizes
harmonics of $f_0$ in linear Hz; on a linear-frequency spectrogram its teeth
are equally spaced, whereas the chord convolver (and a chord-centered
constant-$Q$ bank) show an octave-periodic pattern.

\section{Results}

\subsection{Click and Noise Comparison}

\begin{figure}[t]
\centering
\includegraphics[width=\linewidth,keepaspectratio]{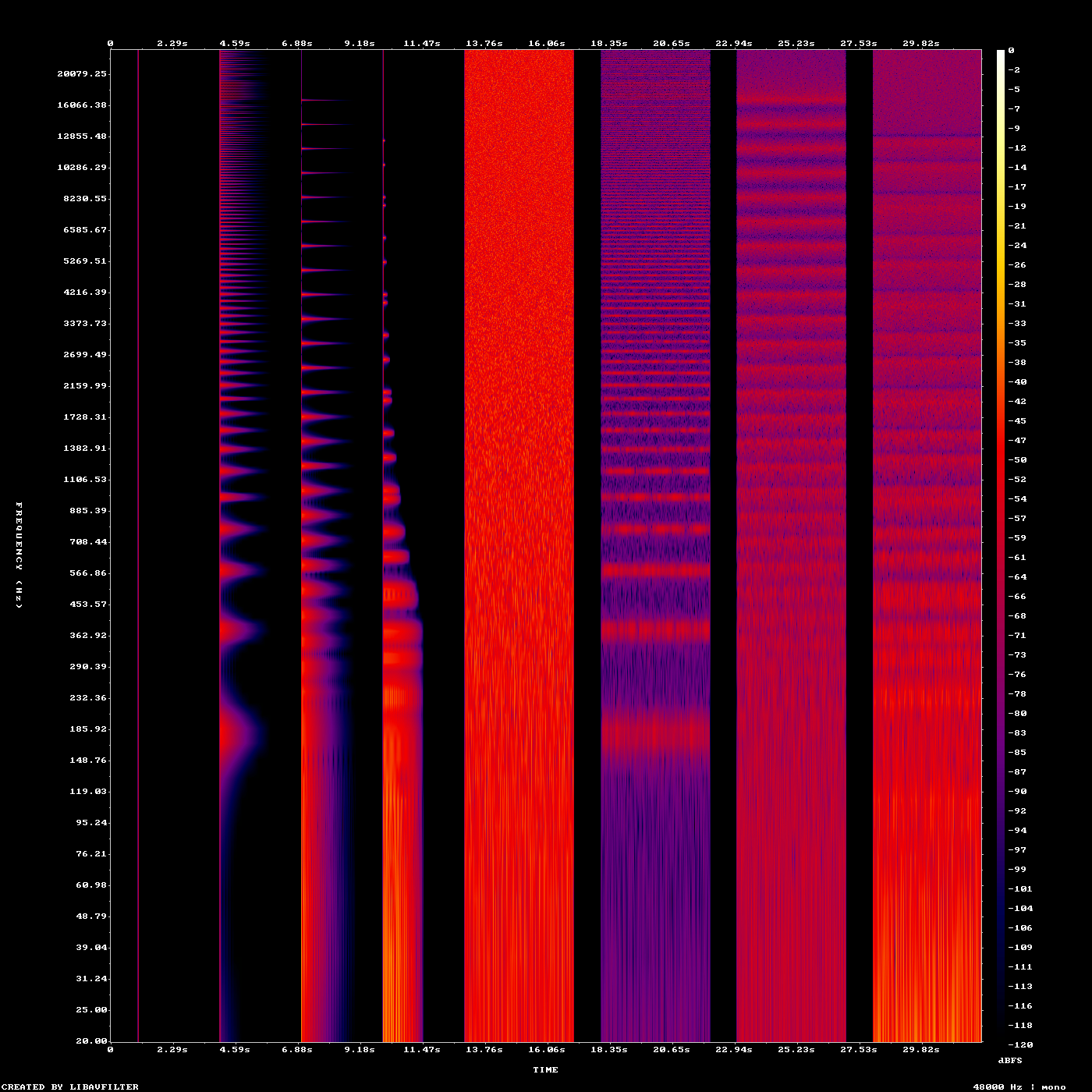}
\caption{Log-frequency spectrogram (left to right): dry click; click comb;
click CQT; click evoke CM7; then the same four treatments of noise.}
\label{fig:spectrograms}
\end{figure}

Figure~\ref{fig:spectrograms} compares treatments of a Dirac-like click and of
white noise: dry source; comb with $f_0=200\,\mathrm{Hz}$; a sparse
constant-$Q$ bank (centers every three semitones); and
\texttt{chord\_convolver} with pitch-classes $0,4,7,11$ (C--E--G--B,
CM7). Segments are loudness-matched for listening, to within
$1.1\,\mathrm{dB}$ A-weighted within each source group. The comb shows
equal spacing in linear Hz; the constant-$Q$ bank shows log spacing without
chord restriction; the convolver shows a sparser chordal lattice, which under
equal-energy normalization stays articulated to the top of the audible range
rather than fading out above the low mids.

\subsection{Sound Walk}

At twilight I walked up from my village in France and alongside the ruins of
Ch\^{a}teau de Lagarde and used a Tascam DR-40 to record my footsteps in fallen
leaves, voices from the village, cars, wind in trees, church bells, and jackdaws. 
A Csound instrument applies
\texttt{chord\_convolver} under score control of onset, duration,
fades, grain duration $T$, wet and dry gains, compensation type, and any
number of pitch-classes.

The first section of this score is (compensation 1 = equal energy;
pitch-classes are MIDI keys reduced modulo 12):

\begin{verbatim}
; instr     onset duration  fadein fadeout      T    wet    dry  comp  pitch-classes
i "evoke"   0.000   29.266    8.00    1.00   0.03   0.30    0.6     1  0  4  7 11 14
i "evoke"  29.266   15.663    1.00    1.00   0.04   0.10    0.6     1  2  5  9 12 14
i "evoke"  44.929   52.171    1.00    1.00   0.06   0.15    0.7     1  5  7  9 14
i "evoke"  97.100   26.208    1.00    1.00   0.05   0.10    0.5     1  2  5  9 12  4
; Three bell strikes
i "evoke" 123.308    1.832    0.05    0.05   0.05   0.10    0.9     1  7 10 13 15
i "evoke" 125.140    1.782    0.05    0.05   0.05   0.20    0.9     1  3  6 10 13 15
i "evoke" 126.922   22.078    0.05   10.00   0.02   0.10    0.8     1  0  4  7 11 14
\end{verbatim}

Effects can be subtle or obvious. Instructive moments include a chord progression within voices,
bells with a subtle effect, bells again with a stronger effect and voices 
following, and ``Bonjour'' with voices following. Audio and further materials are in the
repository~\cite{evoking_harmony_sound_walk,gogins_github_io}.

\section{Discussion}

Evoking harmony is related to comb filtering and to constant-$Q$ analysis, but
it is not the same as either. The lattice is logarithmic and chord-specific, and
the preferred realization is causal grains of short to medium duration plus a
dry Dirac in one impulse response. It works best when harmonic sensation remains
subordinate to the source timbre---a chordal coloration in broadband or mixed
field sound. Wetter settings and pitched sources can produce a clearer
``harmonic shadow,'' ringing, or even increasingly abstract rhythmic and harmonic
textures. Parameter choices ($T$, wet and dry gains, compensation type,
pitch-classes, A-weight shelf) are all audible, and all musical.

The levelling of Section~\ref{sec:levelling} is a compositional choice as well
as a technical one. Equalizing amplitude, energy, or window integral spans
roughly $17\,\mathrm{dB}$ of spectral tilt across the lattice, from a dark
balance in which the upper octaves vanish to a glassy one in which they
dominate. The recordings here use equal energy, so that each pitch-class is
equally present. Equalizing critical bands rather than partials is a further,
audibly distinct option.

\bibliographystyle{unsrt}
\bibliography{icsc2026_evoking_harmony_v2}

\end{document}